\documentclass[superscriptaddress,showkeys]{revtex4-2}
\usepackage{graphicx}
\usepackage{multirow}
\usepackage{placeins}
\usepackage{dcolumn}
\usepackage{bm}
\usepackage[colorlinks=true, citecolor=red, urlcolor = violet, linkcolor= purple, bookmarks=true]{hyperref}
\usepackage{float}
\usepackage[utf8]{inputenc}
\usepackage{amsmath,amsfonts,amssymb}
\usepackage{natbib}
\usepackage{array}
\usepackage{hyperref}
\usepackage{breakurl}
\usepackage{xcolor,verbatim,multirow}
\usepackage[normalem]{ulem}
\usepackage{subcaption} 
\usepackage{wrapfig}
\usepackage{adjustbox}
\usepackage{caption}
\usepackage{mathtools}
\usepackage{pgfplots}
\usepackage{mathrsfs}
\usepackage{booktabs}
\usepackage{orcidlink}
\begin{document} 
\title[]{The Influence of Uniform Magnetic Fields on Strong Field Gravitational Lensing by Kerr Black Holes}
\author{Amnish Vachher\texorpdfstring{\href{https://orcid.org/0009-0003-4991-0662}{\orcidlink{0009-0003-4991-0662}}{}}}\email{amnishvachher22@gmail.com} 
\affiliation{Centre for Theoretical Physics, 
	Jamia Millia Islamia, New Delhi 110025, India}
\author{Arun Kumar\texorpdfstring{\href{https://orcid.org/0000-0001-8461-5368}{\orcidlink{0000-0001-8461-5368}}{}}}\email{arunbidhan@gmail.com}
\affiliation{Institute for Theoretical Physics and Cosmology, Zhejiang University of Technology, Hangzhou 310023, China}
\author{Sushant~G.~Ghosh \texorpdfstring{\href{https://orcid.org/0000-0002-0835-3690}{\orcidlink{0000-0002-0835-3690}}{}}}\email{sghosh2@jmi.ac.in, sgghosh@gmail.com}
\affiliation{Centre for Theoretical Physics, 
	Jamia Millia Islamia, New Delhi 110025, India}
\affiliation{Astrophysics and Cosmology Research Unit, 
	School of Mathematics, Statistics and Computer Science, 
	University of KwaZulu-Natal, Private Bag 54001, Durban 4000, South Africa}
 
\begin{abstract}
We examine gravitational lensing strong deflection limit (SDL) by magnetized Kerr black holes (MKBHs), which are accurate Kerr-Bertotti-Robinson solutions for Kerr black holes in a uniform magnetic field with additional magnetic field strength $B$ apart from mass $m$ and spin $a$. Unlike Kerr-Melvin spacetimes, the MKBH geometry is Petrov type D, without conical singularities, allowing photons to reach asymptotic infinity and making the concept astrophysically feasible. We use the SDL formalism to compute the photon sphere radius, critical impact parameter, deflection angle, and lensing observables, including the image position $\theta_\infty$, angular separation $s$, and relative magnification $r_{\rm mag}$, as well as their relationships with the parameters $a$ and $B$. Our results reveal that the relativistic image's photon sphere and angular size increase with $B$ for prograde orbits as well as retrograde orbits, whereas lensing observables deviate significantly from the Kerr scenario. For M87*, the angular position of relativistic images ranges from $8.67~\mu$as to $26.58~\mu$as for $B=0.1$, and the time delay between the first two images increases from $158.5$~h to $159.6$~h  for prograde orbits, while decreases from $380.8$~h to $340.8$~h for retrograde orbits, respectively, at $B=0.1$ and $\mid a \mid=0.9$. Similarly, for Sgr A*, the image position has a range $11.47~\mu$as to $35.38~\mu$as for $B=0.1$, with time delays enhanced by approximately $0.04$~min for prograde orbits, while a reduction in time delay of $1.57$~min for retrograde orbits, respectively, at $B=0.1$ and $\mid a \mid=0.9$. The relative magnification $r_{\rm mag}$ ranges from $0.926$ to $8.845$ at $B=0.1$. Our outcomes suggest strong gravitational lensing as a powerful tool to investigate the magnetic fields around astrophysical black holes, and in particular, we demonstrate that the MKBH spacetime enables constraints on the parameters $a$ and $B$.
\end{abstract}

\pacs{ } \maketitle
\section{Introduction}
The study of gravitational lensing has a long and fascinating history, commencing with the development of general relativity (GR) itself \cite{Einstein:1936llh}. Over the decades, research on gravitational lensing has dramatically enhanced our understanding of spacetime geometry, singularities, and observational signatures of compact objects \cite{Refsdal:1964yk, Liebes:1964zz, Mellier:1998pk, Bartelmann:1999yn, Perlick:2003vg, Perlick:2004tq, Schmidt:2008hc, Perlick:2010zh, Barnacka:2013lfa}. Darwin \cite{darwin1959gravity} was one of the earliest to apply gravitational lensing to the Schwarzschild black hole, analyzing photon trajectories near the event horizon. Luminet further advanced this study by computing visual images of black holes and deriving what is now called the strong deflection limit (SDL) analytical approximation for light rays passing near the photon sphere \cite{Luminet:1979nyg}. Other significant contributions came from Atkinson, Misner, Thorne, Wheeler, and Ohanian, who explored light propagation and image formation near black holes \cite{atkinson1965light,wheeler2017gravitation, Luminet:1979nyg,ohanian1987black}. The study of strong lensing by black holes in GR was pioneered by \citep{Virbhadra:1999nm}, who examined the lensing properties of Schwarzschild black holes.  Frittelli, Kling, and Newman \cite{Frittelli:1999yf} extended their work, which derived exact lens equations in the form of integral expressions. Later, Bozza \cite{Bozza:2002zj} developed a method for estimating deflection angles in the SDL, which is applicable to different spacetimes.
Subsequent work expanded this analysis to rotating Kerr black holes \citep{Vazquez:2003zm, Bozza:2005tg, Bozza:2006nm}, exposing the dependence of lensing observables on the black hole spin and inclination angle. The SDL is a theoretical framework for investigating lensing near the photon sphere. It was created by \citep{Bozza:2001xd} and later generalized for arbitrary source distances \citep{Bozza:2007gt}. These findings lay the groundwork for linking lensing observables to black hole properties.
 
The recent breakthroughs in observational astrophysics, including the detection of gravitational waves by LIGO/Virgo collaborations and the imaging of the supermassive black holes M87* and Sgr A* by the Event Horizon Telescope (EHT) \cite{EventHorizonTelescope:2019dse, EventHorizonTelescope:2022wkp}, have opened a new era for probing strong gravity regimes and black hole spacetimes. These observations provide crucial opportunities to test the predictions of GR, particularly through the study of gravitational lensing and black hole shadows, and have further intensified interest in this field \citep{Hsieh:2021scb,Ghosh:2020spb,Islam:2021dyk,Islam:2021ful,Kuang:2022xjp,Kumar:2023jgh,Kumar:2020owy,Afrin:2021imp,Kumar:2020hgm,Kumar:2019pjp}.
In modified gravity theories, black hole lensing has special properties that can specify deviations from GR. For example, \citep{Chen:2009eu} explored the gravitational lensing phenomenon in Horava-Lifshitz gravity, whereas \citep{Wei:2014dka} investigated Eddington-inspired Born-Infeld black holes.  Studies of strong gravitational lensing in 4D Einstein–Gauss–Bonnet gravity have been carried out \cite{Islam:2020xmy,Kumar:2020sag}.
 Similarly, the effects of extra dimensions were studied in Kaluza-Klein black holes \citep{Liu:2010wh, Chen:2011ef}, and non-commutative black holes were analyzed by \citep{Ding:2010dc, Ding:2011az}. The role of plasma in lensing was examined by \citep{Liu:2016eju, Bisnovatyi-Kogan:2017kii}, showing how refractive effects alter image positions and magnifications.
Recent work has also focused on black holes in exotic environments, such as those with scalar hair \citep{Zhang:2017vap}, tidal charges \citep{Horvath:2012ru}, or immersed in magnetic fields \citep{Kuang:2022ojj}. The latter is especially essential for astrophysical black holes, as magnetic fields are widespread in active galactic nuclei (AGN) and can considerably alter photon paths \cite{Adler:2022qtb}. The interaction of strong lensing and black hole shadows has been investigated to establish observational restrictions \cite{Kuang:2022xjp,Zhao:2024hep,Walia:2021emv,Kumar:2022fqo,Kumar:2023jgh,Vachher:2024ait,Vachher:2025ytg,Abdujabbarov:2016hnw}. In this study, we investigate the strong gravitational lensing features of Kerr black holes in a uniform magnetic field, a scenario inspired by the magnetized surroundings of AGN. Studies have also explored magnetized black holes as gravitational lenses \cite{Konoplya:2006gg} and analyzed particle motion in the Preston–Poisson spacetime \cite{Konoplya:2006qr}. Further developments include investigations of black hole shadows and lensing in the Melvin universe \cite{Junior:2021dyw} and in Einstein–Maxwell–dilaton spacetimes under strong magnetic fields \cite{Junior:2021svb}.

Among the most well-established exact solutions of GR are the Kerr black holes, representing rotating, uncharged black holes in vacuum \cite{Kerr:1963ud}. While the classic Kerr solution assumes asymptotic flatness, astrophysical black holes are often surrounded by matter and magnetic fields. Indeed, high magnetic fields are expected near compact objects and AGN, which are critical for launching jets and structuring accretion flows. This supports the study of magnetic black holes as more realistic astrophysical models.

Early attempts to include magnetic fields in black hole solutions include the Schwarzschild-Melvin and Kerr-Melvin solutions \cite{Ernst:1976mzr}, derived via Harrison transformations. However, such solutions, while useful, suffer from limitations: they are of Petrov type I, exhibit unbounded ergoregions, and possess pathological asymptotic structures that prevent geodesic motion from escaping to infinity \cite{Gibbons:2013yq}. Recently, a novel exact family of solutions-termed the Kerr-Bertotti-Robinson black holes or Magnetized Kerr black hole (MKBH)-was proposed \cite{Podolsky:2025tle}, describing Kerr black holes embedded in a conformally regular background with the magnetic field strength that is asymptotically uniform and of Petrov type D. These solutions are free from conical singularities, have bounded ergoregions, and allow particles and photons to reach asymptotic regions.
Several works have already explored the physical implications of the MKBH spacetime, including energy extraction via magnetic reconnection \cite{Zeng:2025olq}, geodesic structure and shadow deformation \cite{Wang:2025vsx}, and ergoregion dynamics \cite{Zeng:2025olq}. These findings highlight the astrophysical relevance and distinctiveness of the MKBH spacetime compared to previous magnetized black hole models.

Strong gravitational lensing, in particular, and black hole shadows provide potent probes of spacetime geometry at the event horizon. These observables are directly affected by the features of photon orbits, ergoregions, and magnetic fields. In this work, we study the strong gravitational lensing by the MKBHs. Using the geodesic structure in the MKBH spacetime \cite{Wang:2025vsx}, and applying the techniques of lensing and shadow analysis in the presence of an external magnetic field, we analyze how the magnetic field modifies the photon sphere, lensing observables, and shadow deformation. We compare our theoretical predictions with Kerr black hole predictions. This study complements earlier works on energy extraction \cite{Zeng:2025olq}, photon dynamics \cite{Podolsky:2025tle, Wang:2025vsx,Wang:2025bjf,Zeng:2025tji}, and observational signatures in MKBH spacetimes. It further reinforces the utility of strong field tests of gravity in distinguishing physically viable extensions of the Kerr solution.
We must emphasise that the metric used in this work is a simplified model and does not represent a realistic astrophysical magnetised black hole. Our aim is to study the qualitative influence of the parameter $B$ on the lensing properties and observables. The observables presented here should therefore be regarded as illustrative within the model, rather than as predictions for real astrophysical black holes.

The paper is organized in the following manner: In Section \ref{sec2}, we consider the Kerr black hole solution with a uniform magnetic field or MKBH solutions and derive the effective metric for the motion of photons. Section \ref{sec3} addresses the photon motion in the strong field regime and the photon sphere. In Section \ref{sec3a}, we address the deflection angle of light in the SDL. Section \ref{sec3b} includes lensing observables and time delay in the SDL. In Section \ref{Sec4}, we numerically compute the strong lensing observables and time delays for MKBHs using the supermassive black holes Sgr A* and M87* as the lens and compare them with the Kerr black hole's predictions. In Section \ref{Sec5}, we present a summary of our results and implications for future observations.

In what follows, we set $G = c = 1$ unless otherwise stated and work with the signature convention $\{-, +, +, +\}$. 

\section{Kerr Black Hole Solution in a Uniform Magnetic Field }\label{sec2}
In this section, we review the axisymmetric stationary MKBH solution \cite{Podolsky:2025tle}. The line element for the MKBH black hole in Boyer-Lindquist coordinates can be expressed as
\begin{eqnarray}
	ds^2 = 
 \frac{1}{\Omega^2} \left( g_{tt} \, dt^2 + g_{rr} \, dr^2 + g_{\theta\theta} \, d\theta^2 
	+ g_{\phi\phi} \, d\phi^2 + 2 g_{t\phi} \, dt\, d\phi \right).
	\label{metric}
\end{eqnarray}
where the components are:
\begin{align}
g_{tt} &= -\frac{Q}{\Sigma} + \frac{a^2 \sin^4\theta \, P}{\Sigma},\qquad g_{t\phi} = -\frac{a \sin^2\theta}{\Sigma} \left[ Q - (r^2 + a^2) P \right],\nonumber\\
g_{rr} &= \frac{\Sigma}{Q}, \qquad g_{\theta\theta} = \frac{\Sigma}{P}\qquad g_{\phi\phi} = \frac{\sin^2\theta}{\Sigma} \left[ (r^2 + a^2)^2 P - a^2 \sin^2\theta Q \right].
\end{align}
The metric functions are given by  
\begin{align}
\Sigma   &= r^2 + a^2 \cos^2\theta\,,\qquad P = 1 + B^2 \left( m^2\,\dfrac{I_2}{I_1^2} - a^2 \right) \cos^2\theta\,, \nonumber\\[0mm]
Q &= \left(1 + B^2 r^2 \right) \Delta\,,\qquad \Omega^2 = \left(1 + B^2 r^2 \right) - B^2 \Delta \cos^2\theta\,, \nonumber\\[2mm]
\Delta &= \left(1 - B^2 m^2 \dfrac{I_2}{I_1^2}\right) r^2 - 2m\,\dfrac{I_2}{I_1}\, r + a^2\,, \label{Delta}
\end{align}  
with  
\begin{align}
    I_1 = 1 - \tfrac{1}{2} B^2 a^2\,, \qquad
    I_2 = 1 - B^2 a^2\,. \label{I1I2}
\end{align}  
The MKBH spacetime depends on three parameters: the mass $m$, spin $a$, and magnetic field strength $B$. The spacetime given by Eq.~\eqref{metric} satisfies the Einstein-Maxwell equations, where the 1-form gauge potential is given by (only the real part contributes) \cite{Wang:2025vsx,Podolsky:2025tle}:  
\begin{align}
A_{\mu} \, x^{\mu} = \dfrac{{\rm e}^{\mathrm{i} \gamma}}{2B} \left[ \partial_r\Omega \, \dfrac{a \,  t - (r^2 + a^2) \, \varphi}{r + \mathrm{i} a \cos\theta} + \dfrac{\mathrm{i} \, \partial_\theta\Omega}{\sin\theta} \, \dfrac{ t - a \sin^2\theta \,\varphi}{r + \mathrm{i} a \cos\theta} + (\Omega - 1) \, \varphi \right],  
\label{Amu}
\end{align}  
Here, $\gamma$ is the duality rotation parameter that governs the nature of the gauge potential. For $\gamma=0$, it corresponds to a purely magnetic field, while for $\gamma=\pi/2$, it yields a strictly electric field contribution. The choice of $\gamma$ does not influence the geometry of the spacetime. Notice that the gauge potential vanishes in the limit $B = 0$. 

To find the radii of the horizons, we solve $\Delta=0$
\begin{equation}
r_\pm = \frac{m I_2 \pm \sqrt{m^2 I_2 - a^2 I_1^2 }}{I_1^2 - B^2 m^2 I_2}I_1\,,
\end{equation}  
The roots of this correspond to the locations of the outer and inner horizons of the black hole. The extremal MKBH with degenerate horizons can be obtained, where
\begin{equation}
    m^2I_2-a^2I_1^2=0\label{extermal}\end{equation}which gives
    \begin{equation}
        B^2 =\frac{2}{a^4} \left( m - \sqrt{m^2 - a^2} \right) \sqrt{m^2 - a^2}
    \end{equation}
From this expression, we get the extreme magnetic field. So in the presence of a non-vanishing external magnetic field, the spin parameter $a$ of an extremal MKBH black hole is strictly less than its mass parameter $m$. 
\begin{figure}
    \centering
    \includegraphics[width=0.58\linewidth]{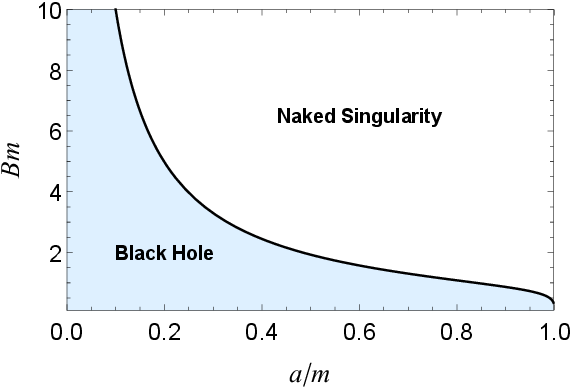}
    \caption{Parameter space for the MKBH spacetime, plotted with the spin parameter $a$ against the magnetic field strength $B$). The solid black curve represents an extremal black hole, which separates the region containing a black hole with two horizons (light blue area below the line) from the region containing a naked singularity (area above the line).}
    \label{fig:parameter}
\end{figure}
For the region $m^2I_2-a^2I_1^2>0$ indicated by the light blue region in Fig.~\ref{fig:parameter}, MKBH has real horizons. The black solid curve, $m^2I_2-a^2I_1^2=0$, divides the region of the black hole and the region of the naked singularity. An important property of MKBH geometry is that it is free from conical singularities along the symmetry axis. In contrast to the $C$-metric, which generically requires conical defects to balance acceleration, the MKBH solution admits a regular axis when constructed via the Harrison transformation, ensuring the correct $2\pi$-periodicity of the azimuthal Killing vector~\cite{Ernst:1976mzr,Griffiths:2009dfa,Astorino:2015lca}. The MKBH spacetime is not asymptotically flat and depicts a Kerr black hole in a uniform magnetic field. The direct product geometry $AdS_{2} \times S^{2}$ supported by a uniform electromagnetic field~\cite{Bertotti:1959pf,Robinson:1959ev} instead approaches the Bertotti--Robinson universe. The external field parameter $B$ controls the conformal structure of the MKBH metric, and the metric reduces to the asymptotically flat Kerr solution~\cite{Kerr:1963ud} in the limit $B \to 0$. However, the external field contribution dominates the asymptotics for nonzero $B$, resulting in a non-flat conformal infinity. Additionally, the geometry is Petrov type D and not conformally flat, except in the specific case $m=0$, as the Weyl scalar $\Psi_{2}$ is nonvanishing for $m \neq 0$, where the solution reduces exactly to the conformally flat Bertotti--Robinson spacetime~\cite{Griffiths:2009dfa,Astorino:2015lca}.

\section{Gravitational Lensing by MKBH spacetime}\label{sec3}
While Bozza's SDL formalism was first derived in the context of asymptotically flat, spherically symmetric spacetimes \cite{Bozza:2002zj}, it has been demonstrated later that it can be applied to more general contexts, such as axisymmetric geometries. Specifically, Bozza and co-workers generalized the method to rotating black holes \cite{Bozza:2002af,Bozza:2005tg}, showing that the logarithmic divergence of the deflection angle close to the photon sphere is a typical feature of stationary spacetimes. In addition, in some theories of gravitation, photons do not travel along the background metric geodesics directly; rather, they travel along geodesics of an effective metric because of self-interactions or couplings \cite{Eiroa:2005ag}. These situations, too, can be accounted for in the framework of SDL as long as the effective metric on which the motion of photons is constrained is used. In addition, the asymptotic flatness assumption may be relaxed to cover spacetimes asymptotically to conformal flatness, as null geodesics are conformally invariant. It is a generalisation highlighted in \cite{Mukherjee:2006ru} that enables the SDL method for black holes with a cosmological constant and associated geometries. Eiroa \cite{Eiroa:2005ag} first emphasized that SDL techniques remain valid if photons follow geodesics of an effective metric rather than the background one, while Mukherjee and Majumdar \cite{Mukherjee:2006ru} extended the method to spacetimes that are asymptotically conformal to flat, thereby establishing its applicability to black holes with a cosmological constant.

The MKBH belongs naturally to this broader class: it is asymptotically uniform to a uniform magnetic background and, unlike the Kerr–Melvin solution, it is free from conical singularities and unbounded ergoregions \cite{Podolsky:2025tle}. Importantly, the MKBH spacetime admits only two horizons, so an observer can always be consistently placed at arbitrarily large distances. Therefore, because the MKBH is asymptotically uniform, free of pathological singularities, and its photon sphere structure produces the same logarithmic divergence, the SDL formalism can be applicable. Hence, we shall follow Bozza's \cite{Bozza:2007gt} recipe for an arbitrary observer, on gravitational lensing in the SDL for MKBH spacetime.

The light rays (null geodesics) do not change under conformal rescaling; the overall conformal factor in the MKBH metric does not affect how photons move. In other words, when we write the metric as  
\begin{equation}
	ds^{2}=\Omega^{-2}\,\tilde g_{\mu\nu}\,dx^{\mu}dx^{\nu}\equiv \Omega^{-2}\mathrm{d\tilde{s}^2}.
\end{equation}
The factor $\Omega^{-2}$ cancels out for light paths, and the motion of photons 
depends only on the metric $\tilde g_{\mu\nu}$. We therefore proceed by working with the conformally related metric $\tilde{g}_{\mu\nu}$ (i.e.~ignoring $\Omega^{-2}$ for geodesics), restricting first to the equatorial plane $\theta=\pi/2$ where the problem is the simplest, and then applying Bozza’s method \cite{Bozza:2005tg,Bozza:2006nm}. The general expressions for strong lensing quantities- such as the photon sphere, the logarithmic divergence of the bending angle, and the expansion in the SDL formalism are expected to remain exactly the same. What changes are the effective metric components $\tilde g_{tt}, \tilde g_{t\phi}, \tilde g_{\phi\phi}, \tilde g_{rr}$, which now include the influence of spin $a$ and the magnetic field strength $B$. These changes alter the photon-sphere radius, the critical impact parameter, and the SDL coefficients ($\bar {a}, \bar {b} $), resulting in distinct magnetic-field signatures in the strong lensing observables.

\begin{figure*}[t!] 
	\begin{centering}
		\begin{tabular}{c c}
		    \includegraphics[scale=0.68]{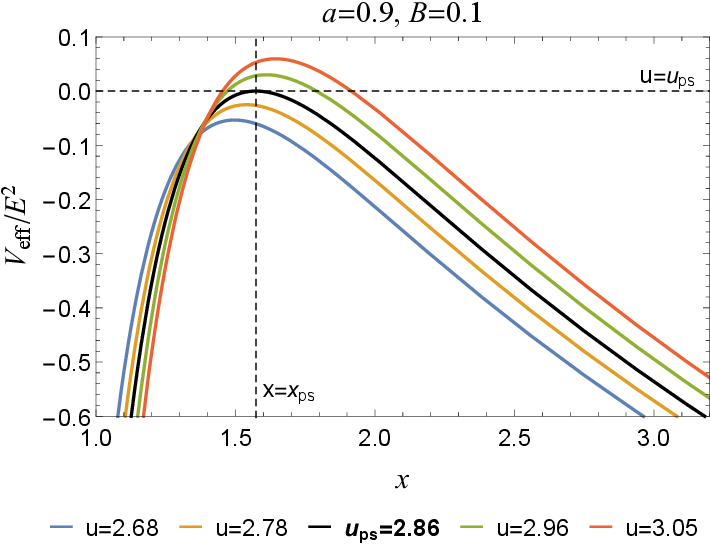}\hspace{0.5cm}
		    \includegraphics[scale=0.68]{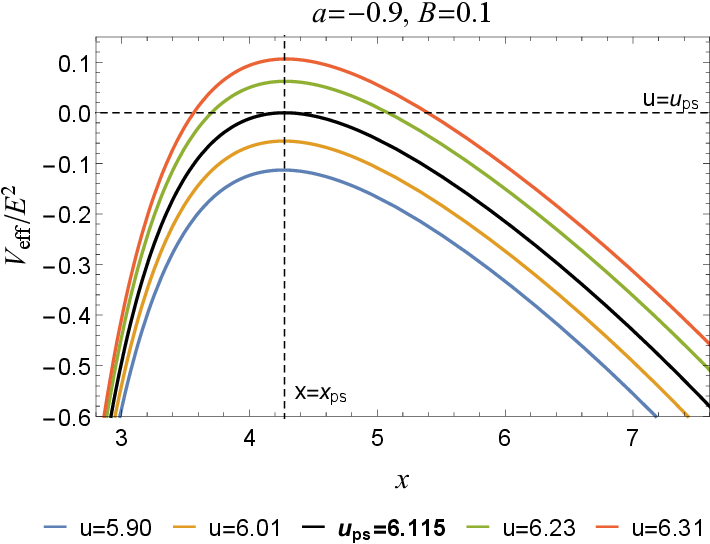}\\
            \includegraphics[scale=0.68]{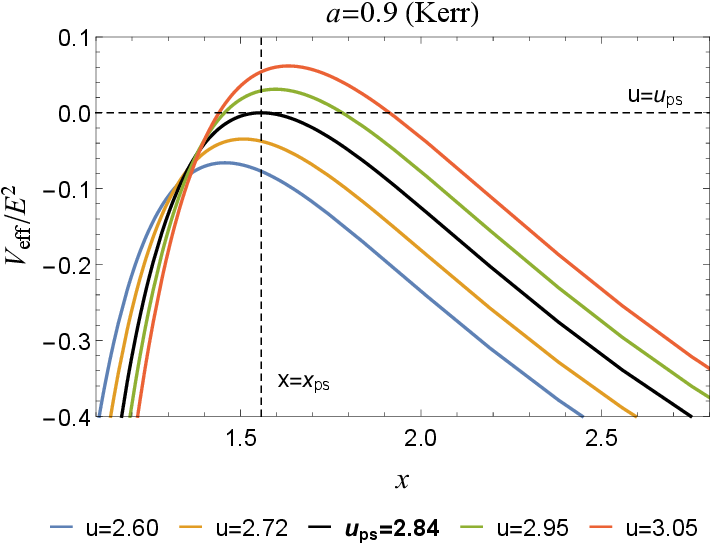}\hspace{0.5cm}
		    \includegraphics[scale=0.68]{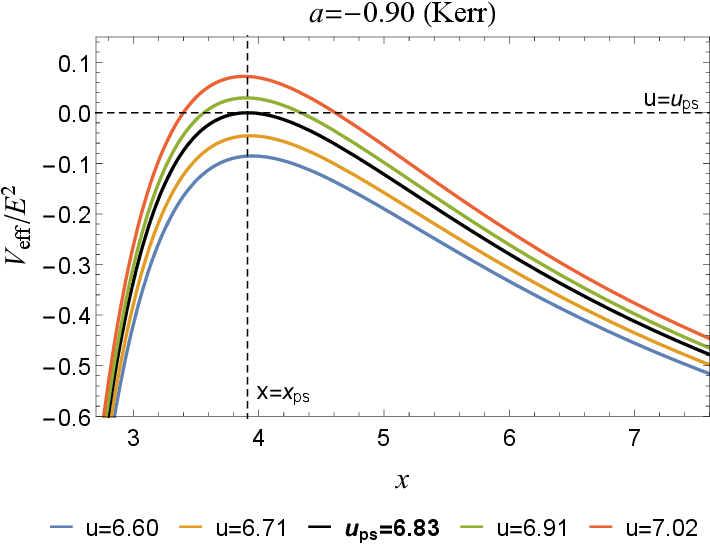}
			\end{tabular}
	\end{centering}
	\caption{ The radial effective potential $V_{\textrm{eff}}$ for photons is plotted against the radial distance $x$ for MKBHs (Upper Panel) and Kerr black holes (Lower Panel) with different values of the impact parameter $u$ and fixed magnetic field $B=0.1$. We keep spin $a=0.9$ (Left Panel) and $a=-0.9$ (Right Panel). The black-curve is for the critical impact parameter $u = u_\text{ps}$, or for an unstable photon orbit with the photon sphere radius $x = x_\text{ps}$. Photons with $u > u_\text{ps}$ (above the black curve) are deflected and scattered to infinity, creating lensed images. Photons with $u < u_\text{ps}$ (below the black curve) fall into the black hole horizon.}\label{plot2}
\end{figure*}

Rescaling all the quantities $r,\, a,\, B$,  and $t$ in units of $m$  and substituting it with \cite{Bozza:2002zj}
\[
r/m \to x, ~~~~a/m \to a,~~~~Bm=B  ~~~~\text{and} ~~~t/m \to t, 
\]
Hence, the rescaled MKBH metric (\ref{metric}) projected onto the equatorial plane can be written as
\begin{eqnarray}\label{NSR}
\mathrm{d\tilde{s}^2}=-g_{tt}dt^2+g_{xx} dx^2 +g_{\phi \phi}d\phi^2-2g_{t\phi}dt\,d\phi
\end{eqnarray}
where
\begin{align}\label{compo}
g_{tt} &= \frac{Q-a^2P}{\Sigma},\qquad g_{t\phi} =\frac{a }{\Sigma} \left[ Q - (x^2 + a^2) P \right], \nonumber \\
g_{xx} &= \frac{\Sigma}{Q},\qquad g_{\phi\phi} = \frac{1}{\Sigma} \left[ (x^2 + a^2)^2 P - a^2 Q \right].
\end{align}
with ${\Sigma} = x^2 $, $P=1$, $Q=(1+B^2x^2)\Delta$ and $\Delta=(1-B^2I_2/I^2_1)x^2-2xI_2/I_1+a^2$. 
The Lagrangian of a particle moving in MKBH spacetime can be written as
\begin{equation}
  \mathcal{L} =\frac{1}{2}g_{\mu\nu}\dot x^{\mu}\dot x^{\nu} 
\end{equation}
where the dot represents the derivative with respect to the affine parameter $\lambda$. From the metric (\ref{NSR}), we have two Killing vectors associated with time translation $\partial_t$ and rotational invariance $\partial_\phi$, and the two conserved quantities \cite{Bozza:2002af}. 
\begin{equation}
E=-p_t=g_{tt}\dot t+g_{t\phi}\dot \phi\qquad\text{and}\qquad L=p_\phi=-g_{\phi t}\dot t+g_{\phi\phi}\dot \phi.
\label{conserve}
\end{equation}
 From Eq.~\eqref{conserve}, we get the two geodesic equations as 
\begin{align}\label{compo}
\dot{t}&=\frac{E(g_{\phi \phi}-ug_{t\phi})}{g_{tt}g_{\phi\phi}+g_{t\phi}^2},~~~~~~~~~~~~
 \dot{\phi}=\frac{E(g_{t \phi}+ug_{tt})}{g_{tt}g_{\phi\phi}+g_{t\phi}^2}
 \end{align}
The above equation, combined with the null condition of the photon ($ds^2=0$), gives the radial geodesic equation as
 \begin{align}
\dot{x}^2+ E^2\left(\frac{u^2g_{tt}+2ug_{t\phi}-g_{\phi \phi}}{g_{xx}(g_{tt}g_{\phi\phi}+g_{t\phi}^2)}\right)=0.
\end{align}
Here, $u=L/E$ \cite{Bozza:2002zj} is the impact parameter of the photon around the black hole. The impact parameter $u$ is the perpendicular distance of the center of mass of the black hole to the initial direction of the photon, which affects the radial motion and determines the minimum approach distance $x_0$. 
The radial equation of motion can also be written as 
\begin{equation}
    \dot{x}^2+V_{\text{eff}}=0
\end{equation}
where $V_{\text{eff}}$ is the effective radial potential, given as \cite{Ghosh:2020spb}:
\begin{equation}
    \frac{V_{eff}}{E^2}=\frac{(g_{tt}u^2+2g_{t\phi}u-g_{\phi \phi})}{g_{xx}(g_{tt}g_{\phi \phi}+g_{t\phi}^2)}\label{veff}
\end{equation}
Here we get a relationship between the effective potential $V_{\text{eff}}$ and the impact parameter $u$. $V_{\text{eff}}$ outlines the path of various photon orbits around the black hole. From Fig.~\ref{plot2}, the photon moves solely within the range where $\mathbf{V_\text{eff}\leq 0}$. At the distance of the closest approach $x_0$, which represents the turning point where $V_\text{eff}=0$, the impact parameter $u$ is
\begin{equation}
   u(x_0)=\frac{-g_{t\phi}(x_0)+\sqrt{g_{tt}(x_0)g_{\phi \phi}(x_0)+g_{t\phi}(x_0)^2}}{g_{tt}(x_0)}, \label{impact}
\end{equation}

\begin{figure*} 
	\begin{centering}
		\begin{tabular}{c c}
		    \includegraphics[scale=0.67]{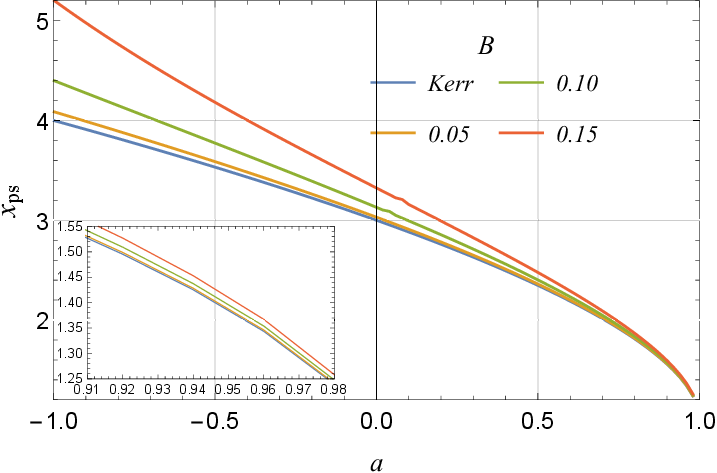}&
		    \includegraphics[scale=0.67]{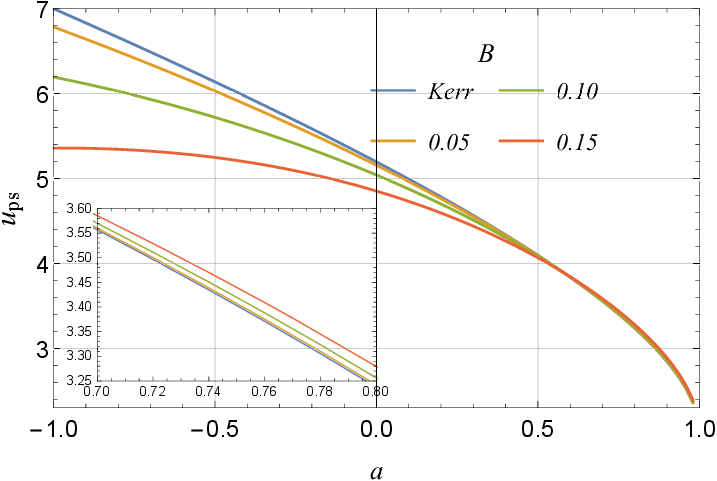}
			\end{tabular}
	\end{centering}
	\caption{Behaviour of the photon orbit parameters with respect to the spin parameter $a$ for different values of the magnetic field strength $B$. (Left) The radius of the photon sphere $x_\text{ps}$ decreases (increases) with increasing spin $\mid a \mid$ for prograde (retrograde) orbits, but increases monotonically with the magnetic field strength $B$. (Right) The critical impact parameter $u_\text{ps}$ shows a similar trend with spin $a$, but it decreases with $B$ for retrograde orbits and prograde orbits with smaller values of $a$, while it increases with $B$ for prograde orbits when the spin parameter is larger.}\label{plot4}
\end{figure*}

Depending on the number of turning points, there are three possible cases: Photons with only two turning point, i.e. $u > u_\text{ps}$;  for photons coming from infinity, only the outer turning point is accessible, so that they reach it and get scattered off to infinity; whereas photons with no turning points, i.e. $u < u_\text{ps}$ are trapped by the black hole. There is also a case where the two turning points merge, that is, $u=u_\text{ps}$, in which photons make circular orbits around the black hole. These circular orbits are unstable, and the condition to determine the radius of these unstable photon orbits $x_\text{ps}$ is given as $V_\text{eff}(x)=V_\text{eff}'(x)=0$ and $V_\text{eff}''<0$. The unstable circular photon orbits $x_\text{ps}$ are occupied by photons, which travel along them until perturbations of small radial size cause these photons either into the black hole or towards infinity. The gravitationally-lensed projection of this photon sphere on the observer's image plane is the black hole shadow. 
 The condition for the solution of an unstable circular photon orbit $x_\text{ps}$ in terms of metric components is given as,

\begin{eqnarray}
\label{ps}
g_{tt}(x)g_{\phi \phi}'(x)-g_{tt}'(x)g_{\phi \phi}(x) + u(g_{tt}'(x)g_{t\phi}(x) - g_{tt}(x)g_{t\phi}'(x)) &=& 0,   
\end{eqnarray}
The unstable circular photon orbit radius is the positive largest root of Eq.~\eqref{ps} \cite{Claudel:2000yi,Virbhadra:2002ju}.
At an unstable photon circular orbit $x_\text{ps}$, the deflection angle of the photon around the black hole becomes divergent.
As the closest approach distance approaches $x_0\to x_\text{ps}$, the impact parameter $u$ is referred to as the critical impact parameter $u_\text{ps}$. Fixing the winding of the light rays to be counterclockwise, we set $a > 0$ (prograde orbits) if the black hole is also rotating in the counterclockwise direction, and $a < 0$ (retrograde) if the black hole turns in the reverse direction of photon twisting, i.e., the clockwise direction.
Figure~\ref{plot4} illustrates a reduction (growth) in the radius of the photon sphere $x_\text{ps}$ and the critical impact parameter $u_\text{ps}$ with the spin parameter $\mid a \mid$ for prograde (retrograde) orbits, suggesting photons forming prograde orbits can get closer to the
black hole than the photons forming retrograde orbits. When the behavior of these parameters is analyzed with the magnetic field strength $B$, we find that for retrograde orbits $x_\text{ps}$ increases while $u_\text{ps}$ decreases with $B$ and for prograde orbits $x_\text{ps}$ always increases with $B$ but $u_\text{ps}$ can increase or decrease with $B$ depending upon the value of $a$. This indicates a larger photon sphere for MKBHs than for Kerr black holes; the critical impact parameter for MKBHs can be larger or smaller than for Kerr black holes, depending upon the spin.

\subsection{Deflection angle in Strong Deflection Limit}\label{sec3a}
\begin{figure*}[t] 
	\begin{centering}
		\begin{tabular}{cc}
		    \includegraphics[scale=0.68]{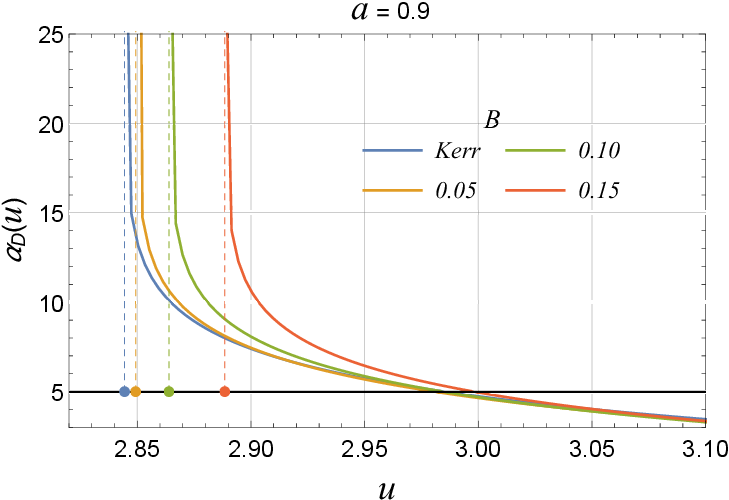}\hspace{0.5cm}
		    \includegraphics[scale=0.68]{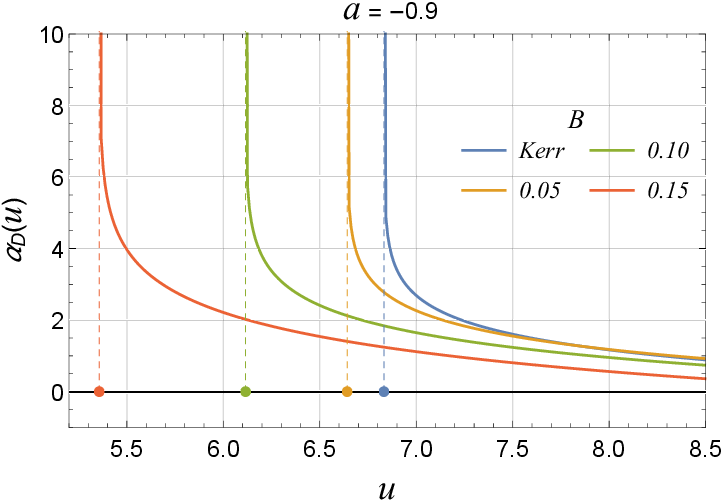}
			\end{tabular}
	\end{centering}
	\caption{The deflection angle $\alpha_D(u)$ of light in the SDL, plotted as a function of the impact parameter $u$ for different values of the magnetic field strength $B$. The left panel is for a spin parameter $a = 0.9$ (prograde) and the right panel for $a = -0.9$ (retrograde). The deflection angle diverges as $u \to u_\text{ps}$ (marked by dots on the horizontal axis). For a specified value of $u$, the deflection angle is smaller (larger) for MKBHs (non-zero $B$) than for a Kerr black hole ($B=0$) for retrograde (prograde) orbits. Furthermore, $u_\text{ps}$ increases with $B$, meaning the divergence happens at a larger impact parameter for prograde orbits compared to Kerr black holes, and $u_\text{ps}$ decreases with $B$ for retrograde orbits.}\label{plot6}
\end{figure*}

In this section, we examine the deflection angle and the coefficients of strong lensing for photons in the vicinity of MKBH spacetime, as well as the influence of its parameters on the lensing coefficients under the strong-field limit. The angle of deflection for the light ray in the equatorial plane at $x_0$ can be computed as: \cite{Islam:2021ful, Islam:2021dyk, Bozza:2002zj, Virbhadra:1998dy}
\begin{eqnarray}\label{bending2}
\alpha_{D}(x_0) &=&  I(x_0) -\pi
\end{eqnarray}
where
\begin{eqnarray}
    &&I(x_0)=2\int_{x_0}^{\infty}\frac{d\phi}{dx}dx
\end{eqnarray}
Substituting the value of $\dot{\phi}$ and $\dot{x}$ from Eq.~\eqref{compo}, we get 
\begin{eqnarray}
     I(x_0)&=&2\int_{x_0}^{\infty}\frac{\sqrt{g_{tt}(x_0) g_{xx}(x) }\left(g_{tt}(x)u+ g_{t\phi}(x)\right)}{\sqrt{g_{tt}(x)g_{\phi\phi}(x)+g_{t\phi}(x)^2}\sqrt{g_{tt}(x_0) g_{\phi\phi}(x)-g_{tt}(x) g_{\phi\phi}(x_0)+2u\left(g_{tt}(x)g_{t\phi}(x_0)-g_{tt}(x_0)g_{t\phi}(x)\right)}}dx,\nonumber\\
\end{eqnarray}
Since the integral is divergent at the photon sphere radius, thus cannot be solved explicitly; therefore, we expand the integral near the unstable photon sphere radius \cite{Virbhadra:1999nm,Claudel:2000yi,Bozza:2002zj}. By defining a new variable $z=1-x_0/x$ \cite{Islam:2022ybr,Zhang:2017vap,Tsukamoto:2016jzh}, we can express an analytic expression of the deflection angle in the SDL for the spacetime (\ref{metric}) as a function of impact parameter $u$ \cite{Bozza:2002zj,Kumar:2020sag,Islam:2020xmy}
\begin{eqnarray}\label{def4}
\alpha_{D}(u) &=& \bar{a} \log\left(\frac{u}{u_\text{ps}} -1\right) + \bar{b} + \mathcal{O}(u-u_\text{ps}),  
\end{eqnarray}  

where $\bar{a}$, $\bar{b}$ are the strong lensing coefficients. Detailed calculations can be found in \cite{Bozza:2002zj,Kumar:2020sag,Islam:2020xmy}. 
Using~\eqref{def4}, we compare and analyze the results of the deflection angle of strong gravitational lensing by MKBH with the Kerr black hole scenario. The effect of the MKBH parameter $B$ for high spin values $a=0.9$ and $a=-0.9$ on the deflection angle of a photon around the black hole as a function of the impact parameter $u$ depicted in Fig.~\ref{plot6}. The deflection angle diverges at larger (smaller) $u_\text{ps}$ as the value of parameter $B$ increases for the prograde (retrograde) orbits. In comparison, we see that the deflection angle for Kerr black holes is larger (smaller) than that of MKBHs for prograde (retrograde) orbits for a fixed value of impact parameter $u$ (c.f. Fig.~\ref{plot6}).

\subsection{Strong Lensing Observables and Time Delay in MKBH Spacetime}\label{sec3b}
\begin{figure}
    \centering
    \includegraphics[width=\linewidth]{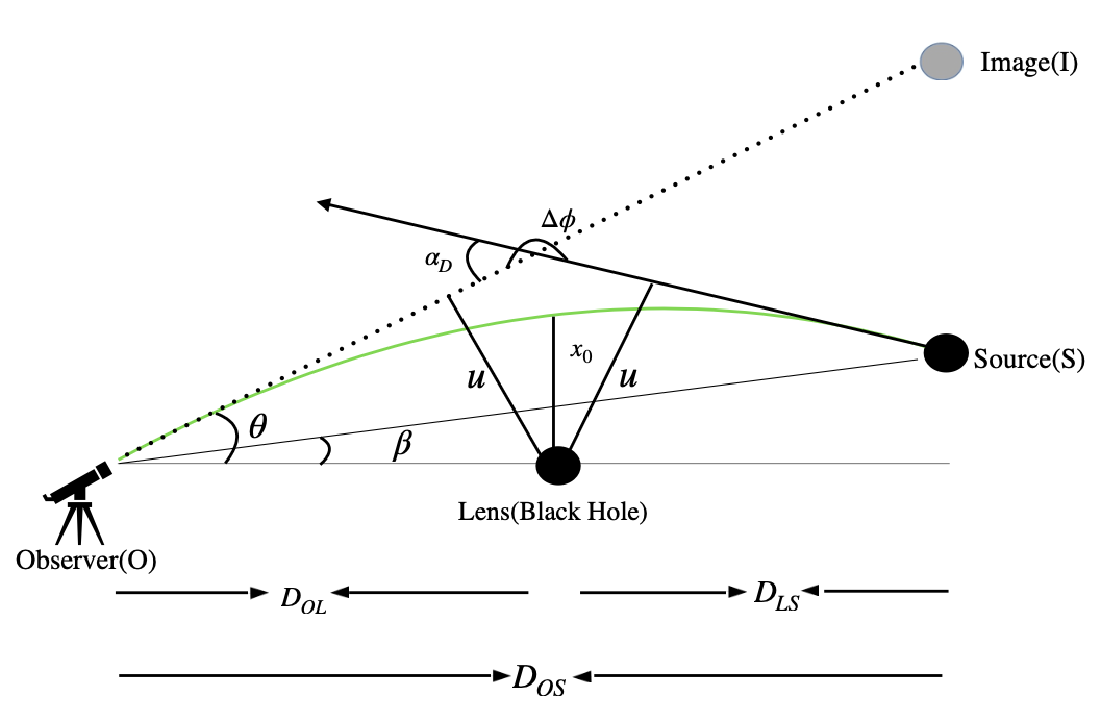}
\caption{A schematic diagram depicting the gravitational lensing. A black hole acts as the lens ($L$), bending the light rays from a distant source ($S$). An observer ($O$) sees multiple, highly magnified, and distorted relativistic images of the source. The angular positions of the source and images relative to the optical axis are given by $\beta$ and $\theta_n$, respectively. The distances between the observer, lens, and source are, respectively, denoted by $D_{OL}$ and $D_{LS}$, and $D_{OS} = D_{OL} + D_{LS}$. }
\label{fig:lensing}
\end{figure}

In this section, we analyze the strong lensing observables calculated by using the lens equation and the deflection angle for a photon. The lens equation is a geometric tool to outline the relations between the observer $O$, the lens $L$, and the light source $S$ (cf. Fig.~\ref{fig:lensing}). Assuming that the source and observer are far from the lens, almost aligned, and placed in a flat spacetime \cite{Bozza:2002zj, Bozza:2008ev, Virbhadra:1999nm}, then the lens equation is written as 
\begin{eqnarray}\label{lenseq}
\beta &=& \theta -\mathcal{D} \Delta\alpha_n,
\end{eqnarray}
where $\Delta \alpha_{n}=\alpha(\theta)-2n\pi$ is the additional deflection angle of a photon moving around $2 n \pi$ and $n$ is an integer $n \in N$, and $0 < \Delta \alpha_n \ll 1$. Moreover, $\beta$ and $\theta$ represent the source and image positions from the optical axis; the distances of the observer(O) and the source(S) to the lens(L) are projected as $D_{\mathrm{LS}}$ and $D_{\mathrm{OS}}$ and $\mathcal{D}$ is the ratio of $D_{LS}$ and $D_{OS}$. It turns out that formulating the lensing equations in terms of the source coordinate $\phi$ is used in recent literature on the strong deflection limit \cite{Bozza:2010xqn,Pietroni:2022cur} . However, we have still chosen to retain the formulation in terms of the angle $\beta$, which represents the source position as seen by the observer, following the robust and widely used technique of Bozza \cite{Bozza:2002zj} invariably applied throughout this paper.
Using the lens Eq.~(\ref{lenseq}) and Eq.~(\ref{def4}), the angular separation between the lens and the $n^{th}$ relativistic image is approximated to~\cite{Bozza:2002af}  
\begin{eqnarray}\label{angpos}
\theta_n &=& \theta_n{^0} +\frac{(\beta-\theta_n{^0})u_\text{ps}e_n}{\bar{a}\mathcal{D} D_{OL}},
\end{eqnarray} 
with 
\begin{eqnarray}
\theta_n{^0} &=& \frac{u_\text{ps}}{D_{OL}}(1+e_n), \label{postion}\\
e_n &=& \text{exp}\left({\frac{\bar{b}}{\bar{a}}-\frac{2n\pi}{\bar{a}}}\right).
\end{eqnarray}
where $\theta_n{^0}$ is the image position at $\alpha=2n\pi$. 
\begin{table}[htb!]
 \begin{centering}
 \setlength{\tabcolsep}{18pt}
 \renewcommand{\arraystretch}{1.2}
	\begin{tabular}{c c c c c c c}
\hline\hline
\multicolumn{2}{c}{}&
\multicolumn{2}{c}{Sgr A*}&
\multicolumn{2}{c}{M87*}\\
{$B$ } & {$a$}& {$\theta_{\infty}$($\mu$as)} & {$s$ ($\mu$as)} & {$\theta_{\infty}$($\mu$as)}  & {$s$ ($\mu$as) } & {$r_{\rm mag}$} \\ \hline
\hline

\multirow{7}{*}{0.0(Kerr)} & -0.85 & 34.1918 & 0.0092695 & 25.6888 & 0.00696431 & 8.60915\\
 & -0.90 & 34.6206 & 0.00871292 & 26.011 & 0.00654615 & 8.6949\\
 & -0.95 & 35.0468 & 0.00819813 & 26.3312 & 0.00615938 & 8.77912\\
 & 0.0 & 26.330 & 0.033 & 19.782 & 0.025 & 6.822\\
 & 0.85 & 15.4769 & 0.337464 & 11.628 & 0.253542 & 3.14933\\
 & 0.90 & 14.4132 & 0.433854 & 10.8289 & 0.325961 & 2.64054\\
 & 0.95 & 13.0845 & 0.57185 & 9.83062 & 0.42964 & 1.93826\\

\hline

\multirow{6}{*}{0.05}
 & -0.85 & 33.2861 & 0.0135531 & 25.0083 & 0.0101826 & 8.30786 \\
 & -0.90 & 33.6559 & 0.0130822 & 25.2862 & 0.00982882 & 8.37419\\ 
 & -0.95 & 34.021 & 0.0126482 & 25.5605 & 0.0095028 & 8.43842 \\
 & 0.85 & 15.5021 & 0.31165 & 11.6469 & 0.234148 & 3.16018 \\
 & 0.90 & 14.4379 & 0.404565 & 10.8474 & 0.303956 & 2.64978\\ 
 & 0.95 & 13.1066 & 0.542236 & 9.8472 & 0.40739 & 1.94478 \\

\hline
\multirow{6}{*}{0.1} & -0.85 & 30.7696 & 0.0283293 & 23.1176 & 0.0212843 & 7.42324\\ 
 & -0.90 & 30.9869 & 0.0286721 & 23.2809 & 0.0215418 & 7.43643\\
 & -0.95 & 31.1952 & 0.0290916 & 23.4374 & 0.0218569 & 7.44628 \\
 & 0.85 & 15.5776 & 0.280862 & 11.7037 & 0.211016 & 3.1926 \\
 & 0.90 & 14.5122 & 0.369644 & 10.9032 & 0.277719 & 2.67751\\ 
 & 0.95 & 13.1731 & 0.507065 & 9.89713 & 0.380966 & 1.96436 \\

\hline

\multirow{6}{*}{0.15} & -0.85 & 27.1191 & 0.0918593 & 20.375 & 0.0690153 & 5.90953\\ 
 & -0.90 & 27.1433 & 0.100307 & 20.3931 & 0.0753619 & 5.83661 \\
 & -0.95 & 27.1556 & 0.110173 & 20.4024 & 0.0827745 & 5.75824 \\
 & 0.85 & 15.7033 & 0.248738 & 11.7981 & 0.18688 & 3.246\\ 
 & 0.90 & 14.6366 & 0.332798 & 10.9967 & 0.250036 & 2.72367 \\
 & 0.95 & 13.2845 & 0.469611 & 9.98086 & 0.352826 & 1.99698 \\                      
\hline
	\end{tabular}
\end{centering}

\caption{The table presents the numerical estimates for the strong lensing observables $\theta_{\infty}$, $s$, and $r_{\rm mag}$ for the supermassive black holes Sgr A* and M87*, modelled as MKBHs. The values are calculated for various combinations of the spin parameter $a$ and magnetic field strength $B$ (both in units of mass $m$). The top section ($B=0$) provides the reference values for a standard Kerr black hole. The table reveals a straightforward trend: $\theta_{\infty}$ increases with $B$, $s$ generally decreases with $B$, and $r_{\rm mag}$ increases with $B$ for high spin for prograde orbits. The trend is opposite for retrograde orbits.}\label{table1}
\end{table}

\begin{table*}[htb!]
 \begin{centering}
 \setlength{\tabcolsep}{18pt}
 \renewcommand{\arraystretch}{1.2}
	\begin{tabular}{ccccccc}
\hline\hline
\multicolumn{2}{c}{}&
\multicolumn{2}{c}{Sgr A*}&
\multicolumn{2}{c}{M87*}\\
{$B$ } & {$a$}& {$\Delta\theta_{\infty}$($\mu$as)} & {$\Delta s$ ($\mu$as)} & {$\Delta\theta_{\infty}$($\mu$as)}  & {$\Delta s$ ($\mu$as) } & {$\Delta r_{mag}$} \\ \hline
\hline
\multirow{6}{*}{0.05}& -0.90&-0.964713& 0.00436924& -0.724803& 0.00328267& -0.320708 \\
& -0.94 & -1.0134& 0.00443427& -0.761381& 0.00333153& -0.336659\\ 
& -0.96 & -1.03823& 0.00446573& -0.780039& 0.00335517& -0.344773\\

& ~0.90&0.0247& -0.0292886& 0.0185575& -0.022005& 0.00924085\\
& ~0.94 & 0.0239532& -0.0301537& 0.0179964& -0.0226549& 0.00831479\\ 
& ~0.96 & 0.0211876& -0.0284022& 0.0159185& -0.021339&0.0057836\\ 
   
\hline

\multirow{6}{*}{0.10}& -0.90 & -3.63376& 0.0199591& -2.7301& 0.0149956& -1.25847\\
& -0.94 &-3.80748& 0.0207036& -2.86062& 0.0155549& -1.31782\\ 
& -0.96 & -3.89582& 0.0210849& -2.92699& 0.0158414& -1.34793 \\
& ~0.90 & 0.0989828& -0.0642096& 0.0743672& -0.0482416& 0.0369678\\
& ~0.94 &0.0960402& -0.0660811& 0.0721564& -0.0496477& 0.0332868 \\ 
& ~0.96 & 0.0849377& -0.0620438& 0.063815& -0.0466144& 0.0231116 \\  
                 
		\hline
	\end{tabular}
\end{centering}
	
	\caption{ The table illustrates the deviation of the lensing observables for MKBHs that from the Kerr black hole case, defined as $\Delta O = O(\text{MKBH}) - O(\text{Kerr})$, for Sgr A* and M87*. The deviations are shown for different spin values $a$ at fixed magnetic fields $B = 0.05$ and $B = 0.1$. A positive (negative) $\Delta\theta_{\infty}$ confirms that the relativistic images appear at a larger (smaller) angular radius for MKBHs for prograde (retrograde) orbits. The negative (positive) $\Delta s$ means a smaller (larger) separation between the first and other images, and the positive (negative) $\Delta r_{mag}$ shows that the first image of MKBH is relatively brighter (dimmer) compared to the Kerr case for prograde (retrograde) orbits.} \label{tab2}
\end{table*}

Due to gravitational lensing, we get a higher flux in the relativistic images. The magnification, for the $ n$-th image, can be expressed as the ratio of the solid angles subtended by the $n^{th}$ image to that of the source 
\begin{equation}
    \mu_n=\left.\frac{\sin\theta}{\sin\beta}\frac{ d\theta}{d\beta}\right|_{\theta\sim\theta_0}
\end{equation}which can be deduced as \cite{Bozza:2002af,Bozza:2002zj}
\begin{eqnarray}\label{mag}
\mu_n &=&  \frac{1}{\beta} \Bigg[\frac{u_\text{ps}^2e_n(1+e_n)}{\bar{a}D_{OL}^2\mathcal{D}} \Bigg].
\end{eqnarray}
\begin{figure*}
	\begin{centering}
		\begin{tabular}{c c}
		    \includegraphics[scale=0.68]{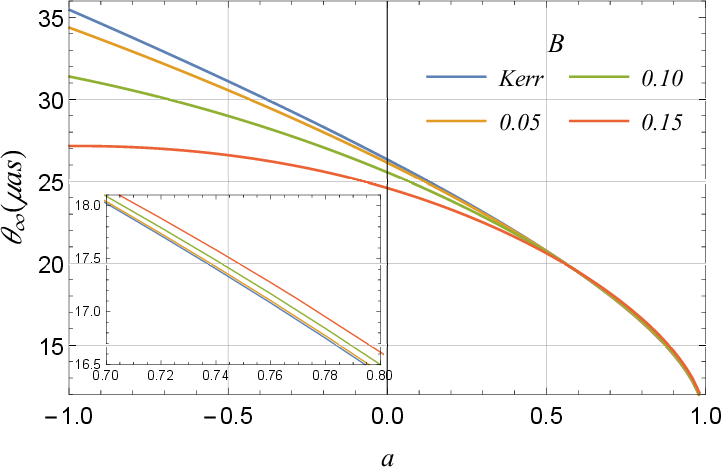}&
            \includegraphics[scale=0.68]{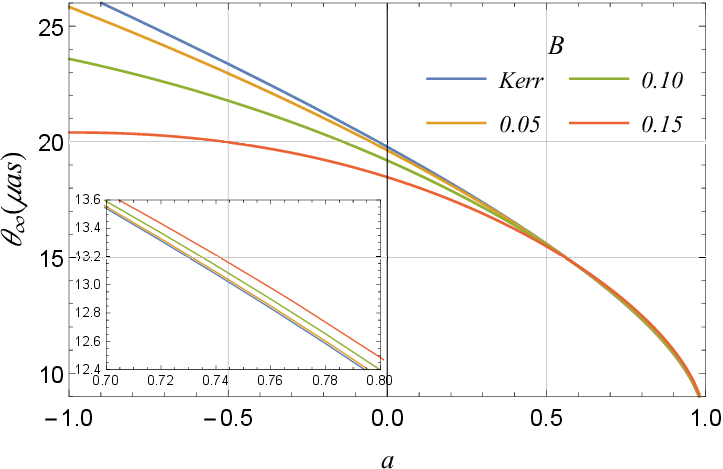}\\
			\includegraphics[scale=0.68]{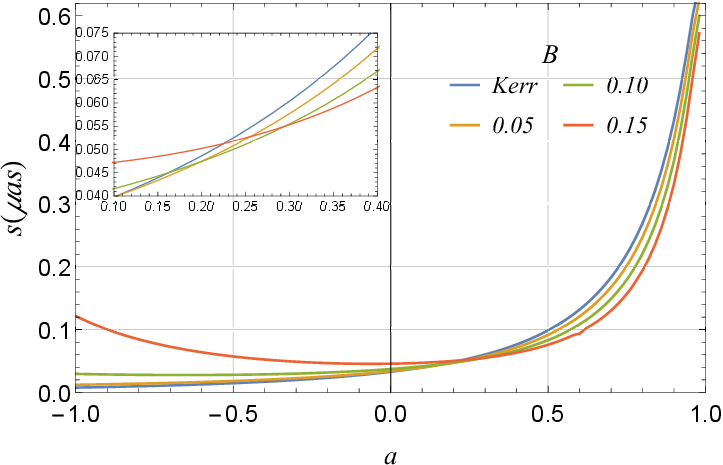}&
			\includegraphics[scale=0.68]{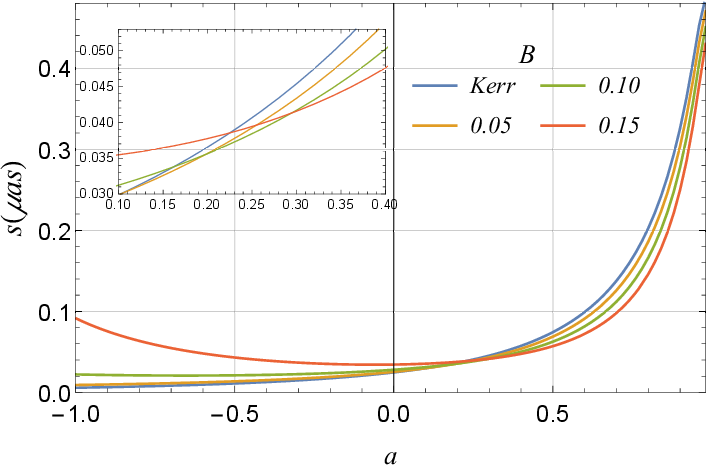}\\
			\end{tabular}
	\end{centering}
	\caption{The behaviour of strong lensing observables $\theta_{\infty}$  and $s$ as a function of spin $a$ for different values of the magnetic field strength $B$. The left panel shows predictions for the supermassive black hole Sgr A*, and the right panel for M87*. This demonstrates how lensing observables can break the degeneracy between the black hole's spin and the surrounding magnetic field strength.}\label{plotobservable}		
\end{figure*}

\begin{figure*}
	\begin{centering}
		\begin{tabular}{cc}
		    \includegraphics[scale=0.68]{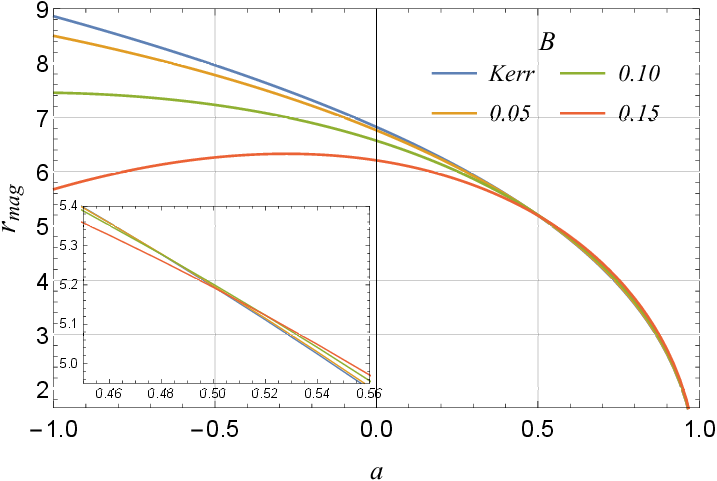}
			\end{tabular}
	\end{centering}
	\caption{The relative flux ratio $r_{\rm mag}$ (the ratio of the magnification of the first image to the sum of all others) plotted as a function of the magnetic field strength $B$ for different spin values $a$. Unlike $\theta_{\infty}$ and $s$, this observable is independent of the black hole's mass and distance. The value of $r_{\rm mag}$ decreases with increasing (decreasing) spin $a$ for prograde (retrograde) orbits but increases or decreases with the magnetic field strength $B$, depending on the value of the  spin parameter.}\label{plotrmag}
\end{figure*}

The brightness of the relativistic images is very faint, since $\mu_n$ is also inversely proportional to $D_{OL}^2$, being a huge value, and further decreases with the order of the image $n$, resulting in images of higher order becoming less visible. We can see that the brightness of the first image dominates over all the other relativistic images. As a result, we consider the case where $\theta_{1}$ can be separated as an outermost single image and the remaining images are grouped, representing the asymptotic position as $\theta_{\infty}$. We define three characteristic observational quantities of relativistic images, as the angular position of the relativistic image $\theta_{\infty}$, the angular separation between $\theta_{1}$ and $\theta_{\infty}$ as $s$, and the relative flux of the first image and all the other images as $r_{\rm mag}$ \cite{Bozza:2002zj} as 
\begin{eqnarray}
\theta_\infty &=& \frac{u_\text{ps}}{D_{OL}},\label{theta}\\
s &=& \theta_1-\theta_\infty \approx \theta_\infty ~\text{exp}\left({\frac{\bar{b}}{\bar{a}}-\frac{2\pi}{\bar{a}}}\right),\label{sep}\\
r_{\text{mag}} &=& \frac{\mu_1}{\sum{_{n=2}^\infty}\mu_n } \approx 
\frac{5 \pi}{\bar{a}~\text{log}(10)}\label{mag1}.
\end{eqnarray}  
By calculating the strong observables, we can determine the nature of the MKBH and compare the results with observational data. Note that the observable $r_{\rm mag}$ does not depend on the distance between the lens and the observer $D_{OL}$, as well as the mass of the black hole.

Another crucial observable in the SDL is the time delay between the relativistic images. This delay in time, created by multiple relativistic images, occurs when the deflection angle exceeds 2$\pi$. Consequently, the travel time of light along divergent paths, corresponding to distinct images, is also inherently varied, resulting in a time delay. The time delay between the $p$-th and $q$-th images located on the same side of the lens is \cite{Bozza:2003cp}
\begin{eqnarray}\label{TD1}
\Delta T_{p,q} \approx 2\pi(p-q) \frac{\tilde{a}}{\bar{a}}\equiv 2\pi(p-q)\frac{\tilde{R}(0,x_\text{ps})}{{R}(0,x_\text{ps})}. 
\end{eqnarray} Using the definition of $\tilde{R}(0,x_\text{ps})$ and $R(0,x_\text{ps})$ from \cite{Bozza:2003cp} and Eqs. \eqref{compo}, \eqref{impact}, Eq. \eqref{TD1} becomes
\begin{eqnarray}\label{TD2}
\Delta T_{p,q} \approx 2\pi(p-q) u_\text{ps}. 
\end{eqnarray}
Hence, the time delay for two arbitrary images on the same side of the lens directly depends on the critical impact parameter $u_{\text{ps}}$ of the photon. Accurately measuring the time delay provides a considerable advantage over dimensional measurements of the system's size, allowing for accurate estimation of the black hole's distance. To measure the time delay, the source must be variable, which is not a strict requirement, since variable stars exist in all galaxies.

\section{Analysis of lensing observables by supermassive black holes,  M87* and Sgr A*}\label{Sec4}

\begin{table*}[htb!]
 \begin{centering}
 \setlength{\tabcolsep}{20pt}
 \renewcommand{\arraystretch}{1.2}
	\begin{tabular}{cccccc}
\hline\hline

{$\mathcal{D}$ } & {$\beta$}& {$\theta_{1p}$($\mu$as)} & {$\theta_{2p}$ ($\mu$as)} & {$\theta_{1s}$($\mu$as)}  & {$\theta_{2s}$ ($\mu$as) } \\ \hline
\hline

\multirow{4}{*}{0.01}&0& 11.1809 & 10.9268 & -21.5395 &-19.5205\\
& 1 & 11.181 & 10.9268 & -21.5399& -19.5205 \\
& 100 & 11.1862&10.9273 & -21.5761 & -19.5225\\
& 10000 & 11.7094 &10.9717 & -25.1971 & -19.9943\\
\hline
\multirow{4}{*}{0.1}&0& 11.1809 & 10.9268 & -21.5395 &-19.5205\\
& 1 & 11.181 & 10.9268  & -21.5395 &-19.5205 \\
& 100 & 11.1815 &10.9269& -21.5432 &-19.5209  \\ 
& 10000 & 11.2338 &10.9313& -21.9053 &-19.5678  \\

\hline
\multirow{4}{*}{0.5}&0& 11.1809 & 10.9268 &-21.5395 & -19.5205\\
& 1 & 11.1809 & 10.9268 &-21.5395 & -19.5205 \\
& 100 &11.1811& 10.9268 &-21.5402&-19.5205 \\
& 10000 & 11.1915 &10.9277& -21.6127 &-19.5299  \\

		\hline
	\end{tabular}
\end{centering}
	
	\caption{The table estimates for the angular positions of the first two relativistic images on the primary ($\theta_{1p}, \theta_{2p}$) and secondary ($\theta_{1s}, \theta_{2s}$) sides (in $\mu\text{as}$) for M87* modeled as an MKBH with parameters $\mid a \mid=0.9$ and $B=0.1$. The positions are calculated for different values of the angular source position $\beta$ (in $\mu\text{as}$) and the distance ratio $\mathcal{D} = D_{LS}/D_{OS}$. Negative sign indicates images formed on the other side of the black hole from the source are formed due to retrograde orbits. The image positions are largely insensitive to the source position $\beta$ for small $\beta$ but show a noticeable dependence for very large misalignments ($\beta = 10000 \mu\text{as}$).  }\label{table4}
\end{table*}

\begin{table*}[htb!]
\resizebox{18cm}{!}{
 \begin{centering}	
	\begin{tabular}{c c c c c c c c c c}
\hline\hline

{$\mathcal{D}$ } & {$B$}& {$\mu_{1p}$}& {$\mu_{1s}$} & {$\mu_{2p}$ }& {$\mu_{2s}$} & {$\Delta_{1p}$}& {$\Delta_{1s}$}  & {$\Delta_{2p}$} & {$\Delta_{2s}$} \\ \hline
\hline

\multirow{3}{*}{0.01}&0.05& $1.337\times10^{-42}$ &$-3.09942\times10^{-43}$& $1.13574\times10^{-43}$& $-1.38495\times10^{-46}$ & 1.06899& 0.711247&	1.07616&	0.529415\\

& 0.10 & $1.23765\times10^{-42}$& $-5.55687\times10^{-43}$&$1.0271\times10^{-43}$ &$-5.88651\times10^{-46}$& 1.1548& 0.396707		&1.18999&0.124559 \\

& 0.15 & $1.14016\times10^{-42}$& $-1.34024\times10^{-43}$ &$9.08943\times10^{-44}$ &$-6.17935\times10^{-45}$ & 1.25354& 0.164482&	1.34468&	0.0118656\\

\hline

\multirow{3}{*}{0.1}&0.05& $1.337\times10^{-43}$ &$-3.09942\times10^{-44}$& $1.13574\times10^{-44}$& $-1.38495\times10^{-47}$ & 1.06899& 0.711247&	1.07616&	0.529415\\

& 0.10 & $1.23765\times10^{-43}$& $-5.55687\times10^{-44}$&$1.0271\times10^{-44}$ &$-5.88651\times10^{-47}$& 1.1548& 0.396707		&1.18999&0.124559 \\

& 0.15 & $1.14016\times10^{-43}$& $-1.34024\times10^{-44}$ &$9.08943\times10^{-45}$ &$-6.17935\times10^{-46}$ & 1.25354& 0.164482&	1.34468&	0.0118656\\
\hline

\multirow{3}{*}{0.5}&0.05& $2.67401\times10^{-44}$&$-6.19884\times10^{-45}$  & $2.27147\times10^{-45}$& $-2.76991\times10^{-48}$ & 1.06899&  0.711247&	1.07616&	0.529415\\
& 0.10 & $2.47531\times10^{-44}$& $-1.11137\times10^{-44}$&$2.05419\times10^{-45}$&$-1.1773\times10^{-47}$& 1.1548& 0.396707	&1.18999&0.124559 \\
& 0.15 & $2.28032\times10^{-44}$ & $-2.68048\times10^{-44}$ &$1.81789\times10^{-45}$ &$-1.23587\times10^{-46}$ & 1.25354& 0.164482&	1.34468&	0.0118656\\
                       
		\hline
	\end{tabular}
\end{centering}
}
	\caption{The table presents estimates for the magnification (brightness) of the first two relativistic images for M87* modeled as an MKBH with spin $\mid a \mid=0.9$ and a source at $\beta = 10~ \mu\text{as}$. The magnifications are extremely small ($\sim 10^{-43}$), ensuring that these higher-order images are highly demagnified. The ratio $\Delta_n = \mu_n(\text{Kerr}) / \mu_n(\text{MKBH}) >1 $, showing that the images are brighter for a Kerr black hole than for an MKBH with the same spin, and this difference in brightness (the ratio $\Delta_n$) increases with the magnetic field strength $B$. The magnification scales inversely with the distance ratio $\mathcal{D}$. }\label{table5}
    
\end{table*} 

\begin{table*}[tbh!]
	\begin{ruledtabular}
		\begin{tabular}{c c c c c c c c} 
        \multicolumn{4}{c}{}&
\multicolumn{2}{c}{Prograde($a>0$)}&
\multicolumn{2}{c}{Retrograde($a<0$)}\\
				Galaxy   &           $M( M_{\odot})$      &          $D_{OL}$ (Mpc)   &     $M/D_{OL}$ & $\Delta T^s_{2,1}(\text{Kerr})$&$\Delta T^s_{2,1}(\text{MKBH})$ &$\Delta T^s_{2,1}(\text{Kerr})$&$\Delta T^s_{2,1}(\text{MKBH})$          \\
			\hline
           
			Milky Way& $  \mathbf{4.28\times 10^6}	 $ & $0.0083 $ &       $2.471\times 10^{-11}$ &6.26416& 6.30718& 15.0466& 13.4673    \\
             M87&$ 6.5\times 10^{9} $&$ 16.68 $
&$1.758\times 10^{-11}$& 9513.33& 9578.67& 22851.1& 20452.7\\			
		
			 NGC 4472 &$ 2.54\times 10^{9} $&$ 16.72 $
&$7.246\times 10^{-12}$& 3717.52& 3743.05& 8929.5& 7992.27\\
			
			 NGC 1332 &$ 1.47\times 10^{9} $&$22.66  $
&$3.094\times 10^{-12}$&2151.48& 2166.25& 5167.86& 4625.45 \\
		
			 NGC 4374 &$ 9.25\times 10^{8} $&$ 18.51 $
&$2.383\times 10^{-12}$& 1353.82& 1363.12& 3251.89& 2910.57\\
			
			NGC 1399&$ 8.81\times 10^{8} $&$ 20.85 $
&$2.015\times 10^{-12}$& 1289.42& 1298.28& 3097.2& 2772.12\\
			 
			  NGC 3379 &$ 4.16\times 10^{8} $&$10.70$
&$1.854\times 10^{-12}$& 608.853& 613.035& 1462.47& 1308.97\\
			
			 NGC 4486B &$ 6\times 10^{8} $&$ 16.26 $
&$1.760\times 10^{-12}$ & 878.154& 884.185& 2109.33& 1887.94\\
		
			 NGC 1374 &$ 5.90\times 10^{8} $&$ 19.57 $ &$1.438\times 10^{-12}$&863.518& 869.448& 2074.18& 1856.47\\
			    
			NGC 4649&$ 4.72\times 10^{9} $&$ 16.46 $
&$1.367\times 10^{-12}$& 6908.14& 6955.59& 16593.4& 14851.8\\
		
			NGC 3608 &$  4.65\times 10^{8}  $&$ 22.75  $ &$9.750\times 10^{-13}$& 680.569& 685.243& 1634.73& 1463.15\\
		
			 NGC 3377 &$ 1.78\times 10^{8} $&$ 10.99$
&$7.726\times 10^{-13}$ & 260.519& 262.308& 625.768& 560.088\\
		
			NGC 4697 &$  2.02\times 10^{8}  $&$ 12.54  $ &$7.684\times 10^{-13}$& 295.645& 297.676& 710.142& 635.606\\
			 
			 NGC 5128 &$  5.69\times 10^{7}  $& $3.62   $ &$7.498\times 10^{-13}$& 83.2783& 83.8502& 200.035& 179.039\\
			
			NGC 1316&$  1.69\times 10^{8}  $&$20.95   $ &$3.848\times 10^{-13}$& 247.347& 249.045& 594.128& 531.769\\
			
			 NGC 3607 &$ 1.37\times 10^{8} $&$ 22.65  $ &$2.885\times 10^{-13}$& 200.512& 201.889& 481.631& 431.079\\
			
			NGC 4473 &$  0.90\times 10^{8}  $&$ 15.25  $ &$2.815\times 10^{-13}$& 131.723& 132.628& 316.4& 283.191\\
			
			 NGC 4459 &$ 6.96\times 10^{7} $&$ 16.01  $ &$2.073\times 10^{-13}$& 101.866& 102.565& 244.682& 219.001\\
		
			M32 &$ 2.45\times 10^6$ &$ 0.8057 $
&$1.450\times 10^{-13}$ & 3.5858& 3.61042& 8.6131& 7.70908   \\
		
			 NGC 4486A &$ 1.44\times 10^{7} $&$ 18.36  $ &$3.741\times 10^{-14}$ & 21.0757& 21.2204& 50.624& 45.3105\\
			 
			NGC 4382 &$  1.30\times 10^{7}  $&$ 17.88 $  &$3.468\times 10^{-14}$& 19.0267& 19.1573& 45.7022& 40.9053\\
			CYGNUS A &$  2.66\times 10^{9}  $&$ 242.7 $  &$1.4174\times 10^{-15}$&3893.15& 3919.89& 9351.37& 8369.86\\
		\end{tabular}
	\end{ruledtabular}
\caption{Predicted time delay $\Delta T_{2,1}$ (in hours) between the first and second relativistic images on the same side of the lens for supermassive black holes at the centers of nearby galaxies. The time delay is calculated for two scenarios: a Kerr black hole with $\mid a \mid=0.9$ and an MKBH with $\mid a \mid=0.9, B=0.1$. The mass $M$ (in solar masses) and distance $D_{OL}$ (in Mpc) are given. The time delay is invariably larger for the MKBH case due to its larger critical impact parameter $u_\text{ps}$ for prograde orbits, whereas it exhibits the opposite behaviour for retrograde orbits. For example, the delay for M87* increases from 9513.33 minutes (Kerr) to 9578.67 minutes (MKBH) for prograde orbits, while it decreases from 22851.1 minutes (Kerr) to 20452.7 minutes (MKBH) for retrograde orbits. The masses and the central black hole distances are adopted from \cite{EventHorizonTelescope:2022xqj,EventHorizonTelescope:2019dse,McConnell:2012hz,Kormendy:2013dxa}
	}\label{table3} 
\end{table*}

We model EHT targets, supermassive black holes M87* and Sgr A*, as Kerr black hole in a uniform magnetic field to estimate their lensing observables and compare our results with those for Kerr black holes. Based on the latest astronomical observation data, the estimated mass of M87* is $\left(6.5\pm0.7\right)\times{10}^9M_\odot$, and its distance is $d=16.8$ MPc \cite{EventHorizonTelescope:2019ggy}. The estimated mass of SgrA* is $4.28_{\pm0.10}^{\pm0.21}\times{10}^6M_\odot$ \cite{2017ApJ...837...30G}, and its distance is $d=8.32^{\pm0.07}_{\pm0.14}$ KPc \cite{2017ApJ...837...30G}.

Here, it is important to point out that the conversion formula relating $|B|$ to the physical magnetic field $B_{\rm phys}$ is
 \begin{equation}
 	B_{\rm phys} = \frac{|B|\, c^4}{G^{3/2} M_{\bullet}}.
 \end{equation}
 Taking $M_{\bullet} \approx 4.28 \times 10^6 M_\odot$ for Sgr A* and $M_{\bullet} \approx 6.5 \times 10^9 M_\odot$ for M87*, our chosen range $|B| = 0.05$--$0.15$ corresponds to extremely strong near-horizon fields of roughly
 $B_{\rm phys} \sim 10^{10}\text{--}10^{11}~\mathrm{G\ (Sgr~A^*)}, \quad
 B_{\rm phys} \sim 10^7\text{--}10^8~\mathrm{G\ (M87^*)}$.
 These values are far above the 1--30~G measured by the Event Horizon Telescope (EHT) in the accretion flows of both Sgr A* and M87*  \cite{EventHorizonTelescope:2021srq}.
  We must stress that the metric used is primarily of theoretical interest and cannot be considered as a real astrophysical magnetized black holes. Accordingly, the observables shown in the tables should be understood as illustrative within this model framework, rather than as quantitative predictions for real black holes. Choosing $|B|$ in this way allows us to explore the maximum possible impact of magnetic fields on photon trajectories and lensing effects near the horizon.

We performed a comparative analysis of the relativistic image positions $\theta_\infty$ and lensing observables, $s$ and $r_{mag}$, for MKBHs with their Kerr black hole counterparts and present the results in Table \ref{table1}. The asymptotic angular size $\theta_\infty$ grows when the magnetic field strength parameter $B$ grows, but diminishes as the spin parameter $a$ gets larger in magnitude for prograde orbits, while the asymptotic angular size $\theta_\infty$ decreases with $B$ and increases with $a$ for retrograde orbits. Analogously, image separation $s$ falls (rises) with $B$ ($a$) for prograding photons while rises (falls) with $B$ ($a$) for retrograding photons (cf. Fig.~\ref{plotobservable} and Table \ref{table1}). For M87* and Sgr A*, the deviation of lensing observables with a Kerr black hole is computed in Table~\ref{tab2}. We see that the difference in the lensing observable $\theta_\infty$ grows as the magnetic field strength parameter $B$ gets larger for photons moving in prograde orbits, which implies that the relativistic images produced by MKBHs are bigger than Kerr black hole images, while showing opposite behavior for retrograde orbits. The relative flux $r_{\rm mag}$ increases as the magnetic field strength parameter $B$ increases for prograde orbits (cf. Fig~\ref{plotrmag}), illustrating that the brightness of the first relativistic image in comparison to all the remaining images enhances with the increase of the magnetic field parameter $B$ values. As compared to the Kerr black hole, the MKBHs produce brighter images than the Kerr black hole with the increase of the magnetic field strength $B$ (cf. Table \ref{tab2}). For retrograde orbits, we observe a decrease in the relative flux with $B$ and as the spin parameter $\mid a \mid$ increases, the deviation from Kerr black holes becomes larger for retrograde orbits, producing dimmer images than Kerr black holes for larger spin values.

With Eq.~\eqref{angpos}, we have also contrasted the main and secondary angular position with the angular position of the source ($\beta$) and the observer-lens to source distance ratio $\mathcal{D}$ for a given parameter value $a$ and $B$ in Table.~\ref{table4}. In comparison, it is clear that $\mid\theta_s\mid>\mid\theta_p\mid$, and the relativistic image gets a smaller angular size as the order ($n$) of the angular position increases. Observe that the negative sign in the secondary image indicates images on the other side of the black hole from the source are formed by photons moving in retrograde orbits.

The magnifications of $n$-th order images of MKBHs and their comparison with that of the Kerr black hole for the M87* are given in Tables \ref{table5}. We observed that higher-order images are highly demagnified, and inversely proportional to the distance ratio $\mathcal{D}$ and source position $\beta$. The brightness decreases as the order $n$ of relativistic images increases. The brightness variation of relativistic images becomes noticeably larger as the magnetic field strength $B$ increases for primary images created by prograde orbits. This means that as the magnetic field strength $B$ is increased, the brightness of relativistic images diminishes, with Kerr black holes ($B=0$) producing the brightest images. The magnification exhibits opposite behavior for secondary images formed by photons moving in retrograde. Negative sign indicates images formed on the other side of the black hole from the source.

The delay time between the first image and the second image, $\Delta T_{2,1}$ for supermassive black holes at the center of galaxies nearby, considering them as MKBH, and comparison with a Kerr black hole is given in Table~\ref{table3}. The time delay $\Delta T_{2,1}$ for $22$ galaxies is estimated, for instance, by MKBHs, as M87 * estimates a time delay of $159.64$~ h and $340.87$~h, respectively, for the magnetic field parameter value $B=0.1$ and spin parameter value $\mid a \mid=0.9$ for prograde and retrograde orbits with deviation from the Kerr black hole for the same spin value can reach up to $17.52$~ h and $40.03$~ h, respectively, as photons take more time for photons to arrive when photons are in retrograde motion. Measuring the time delay in Sgr A* is far more challenging and shorter, and we must wait for the next-generation EHT (ngEHT) to do the same.

\section{\label{Sec5}Conclusion}
In this work, we have examined the strong gravitational lensing by MKBHs, which describe Kerr black holes immersed in a uniform external magnetic field. Unlike the  Kerr--Melvin geometry, the MKBH spacetime is free from pathologies such as unbounded ergoregions or conical singularities, while retaining Petrov type D character and permitting photon trajectories to extend to infinity. This makes the model astrophysically viable for describing supermassive black holes in magnetized environments, such as those expected in active galactic nuclei.  

By applying the SDL formalism, we derived the gravitational lensing coefficients, photon sphere radius, and critical impact parameter for MKBHs, and studied their dependency on the spin parameter $a$ and the magnetic field strength $B$. Our outcomes indicate that the photon sphere radius increases monotonically with $B$, both for prograde and retrograde orbits. At the same time, the critical impact parameter shows a more intricate dependence: it decreases with $B$ for retrograde orbits. Still, it may either increase or decrease with $B$ for prograde orbits, depending on the value of spin. The deflection angle exhibits the familiar logarithmic divergence near the photon sphere, but the location of the divergence is shifted by the presence of $B$. The deflection angle diverges for prograde orbits as the impact parameter increases, but the opposite pattern is seen for retrograde orbits. As a result, the external magnetic field is established as an independent parameter that governs the shape of photon orbits and, in turn, the visible lensing signals. 

The strong lensing observables associated with MKBHs also exhibit distinctive behavior. We also show that the angular position of the relativistic image, $\theta_\infty$, increases with $B$ for prograde orbits but decreases for retrograde orbits, while the angular separation $s$ between the outermost image and the image cluster shows the complementary trend. The relative magnification $r_{\rm mag}$ decreases with increasing spin $a$, but increases with $B$ for prograde orbits, whereas for retrograde orbits its behavior is reversed. These dependencies mean the degeneracy between the spin and the magnetic field in shaping the strong lensing observables. An accurate parameter estimation for astrophysical black holes, therefore requires a combination of multiple observables, so that the contributions from $a$ and $B$ can be disentangled.

We further analyzed the time delays between successive relativistic images, which also carry a distinct imprint of the magnetic field. For prograde photon orbits, the time delay grows with $B$, whereas for retrograde photon orbits it decreases. For M87*, the difference relative to the Kerr case may be as large as several tens of hours for $B \sim 0.1$, while for Sgr A* the corresponding deviation is smaller, of order minutes.  Such differences could in principle be probed by future astronomical observation, such as the ngEHT, particularly for sources exhibiting intrinsic variability. Our numerical estimates for M87* and Sgr A* confirm that even relatively small values of $B$ lead to observable modifications of lensing quantities, underlining the astrophysical relevance of the MKBH black holes. 

Taken together, these results highlight that magnetic fields play a nontrivial role in the strong lensing regime of rotating black holes. The MKBH spacetime, being regular and astrophysically motivated, provides a rich theoretical setting to model and interpret the observational data from the EHT and its successor ngEHT. In particular, the combined analysis of shadow size, image separation, flux ratios, and time delays has the potential to place meaningful constraints on both the spin $a$ and the magnetic field strength $B$ in the near-horizon regions of supermassive black holes. Lastly, we must stress that the MKBH metric in this paper is mainly of theoretical interest only and is not intended to describe real astrophysical magnetized black holes.

Future work may extend the present analysis in several directions. An immediate extension is to include off-equatorial photon trajectories, which would provide a more complete picture of lensing in the MKBH black hole background. Incorporating plasma effects, which are expected to be relevant in realistic accretion flows, would further enhance astrophysical applicability. Finally, full ray-tracing simulations of accretion flows and hot spots in MKBH geometry could directly connect the theoretical predictions to VLBI images, improving the discriminating capability of EHT and ngEHT observations. Such expansions would considerably strengthen our ability to test the interplay of gravity and magnetic fields in the strong-field regime, and offer deeper insight into the physical environments of supermassive black holes. 

\section{Acknowledgments} 
S.G.G. gratefully acknowledges the support from ANRF through project No. CRG/2021/005771. This work is supported by the Zhejiang Provincial Natural Science Foundation of China under Grants No.~LR21A050001 and No.~LY20A050002, the National Natural Science Foundation of China under Grants No.~12275238 and No. ~W2433018, the National Key Research and Development Program of China under Grant No. 2020YFC2201503, and the Fundamental Research Funds for the Provincial Universities of Zhejiang in China under Grant No.~RF-A2019015.

\section{Data Availability Statements}
No new data were generated or analysed in support of this research, nor was any third-party data analyzed.
\bibliography{reference}
\end{document}